\documentclass[twocolumn,prd,amsmath,amssymb,showpacs,nofootinbib]{revtex4}

\usepackage{epsfig,verbatim,paralist}

\renewcommand{\vec}[1]{\boldsymbol{\mathrm{#1}}}

\begin{document}

\title{Thermal recoil force, telemetry, and the Pioneer anomaly}

\author{Viktor T. Toth}
\address{Ottawa, Ontario  K1N 9H5, Canada}
\homepage{http://www.vttoth.com/}

\author{Slava G. Turyshev}
\affiliation{Jet Propulsion Laboratory, California Institute of Technology\\
4800 Oak Grove Drive, Pasadena, California 91109-8099, USA}
\email{turyshev@jpl.nasa.gov}

\date{\today}
\begin{abstract}
Precision navigation of spacecraft requires accurate knowledge of small forces, including the recoil force due to anisotropies of thermal radiation emitted by spacecraft systems. We develop a formalism to derive the thermal recoil force from the basic principles of radiative heat exchange and energy-momentum conservation. The thermal power emitted by the spacecraft can be computed from engineering data obtained from flight telemetry, which yields a practical approach to incorporate the thermal recoil force into precision spacecraft navigation. Alternatively, orbit determination can be used to estimate the contribution of the thermal recoil force. We apply this approach to the Pioneer anomaly using a simulated Pioneer 10 Doppler data set.
\end{abstract}



\maketitle

\section{Introduction}

Precision navigation of spacecraft requires a detailed knowledge of small forces affecting its trajectory. This includes the recoil force associated with thermal radiation. It recently became clear that the small force due to the radiation of waste heat from the spacecraft itself cannot be ignored \cite{MDR2005}. This is especially true in the case of the Pioneer~10 and 11 spacecraft. A small, anomalous Doppler residual is present in these spacecrafts' radio signal, and it is believed to be caused by an acceleration of unknown origin \cite{JPL1998,JPL2002,MDR2005}. Investigations of this anomaly indicated that its magnitude is comparable to that of the acceleration due to a thermal recoil force \cite{KATZ1998,MURPHY1998,LKS2003}. Therefore, the possibility that the anomalous acceleration is caused by an anisotropic emission of the heat generated on board of the spacecraft must be investigated \cite{MEXPROC2007}.

Surprisingly, there is no established formalism for incorporating the thermal recoil force into precision spacecraft navigation\footnote{A closely related topic is that of the reemission of solar thermal radiation by asteroids and the resulting recoil force, otherwise known as the Yarkovsky effect \cite{Bottke2006}.}. In fact, most of the literature is dealing with energy distribution processes, but not momentum transfer. As a result, some authors were forced to develop {\em ad hoc} formalisms in order to estimate the effects of the thermal recoil force on spacecraft trajectories \cite{ROS1990,ANT1992,VIG1994,DUH2006}; while others (e.g., \cite{JPL2002}) only roughly estimate the magnitude of this force, but do not precisely determine its direction and temporal dependence, nor do they provide formal error estimates. However, this is exactly what we need to investigate the nature of the Pioneer anomaly.

For navigational purposes, one is interested in the magnitude, direction, and temporal evolution of forces acting on a spacecraft \cite{MG2005}. For the recoil force due to internally generated heat these can be estimated using available knowledge of the spacecraft's geometry, thermal properties, and state. The geometry and thermal properties of the spacecraft (excluding effects of aging) can be established from design documentation, prelaunch test results, and calibration experiments performed after launch. The spacecraft's state, in turn, can be obtained for any given moment of time from its telemetry, which contains unique data on the power consumption, thermal status, and physical configuration of the craft \cite{MDR2005}.

The purpose of this paper is to develop the methods, tools, and procedures that are needed for the evaluation of the thermal recoil force as a likely cause of the anomalous acceleration of the Pioneer~10 and 11 spacecraft. This investigation of the on-board small forces is performed in conjunction with the analysis of the Doppler data, allowing us to show, for the first time, how precision orbit determination can also be used to estimate a spacecraft's thermal properties.

This paper is organized as follows. Our analysis begins with considering heat conduction and radiation in Sec.~\ref{sec:transfer}. We calculate radiation pressure in Sec.~\ref{sec:radpress}. Next, in Sec.~\ref{sec:recoil}, we explore in detail the resulting recoil force and its relationship with the amount of heat generated. In Sec.~\ref{sec:errors}, we analyze the accuracy with which the recoil force can be determined. Next, in Sec.~\ref{sec:orbit}, we connect the computation of the recoil force with orbital analysis. In Sec.~\ref{sec:application}, we apply what we learned to simulated Pioneer~10 orbital data. We present our conclusions in Sec.~\ref{sec:end}.

~\par~\par

\section{Heat conduction and radiation}
\label{sec:transfer}

Heat flows from internal heat sources to external radiating surfaces via three mechanisms: convection, conduction, and radiation. For the purposes of this paper, we ignore convection within the spacecraft\footnote{We note that convection may be significant in the case of spacecraft that utilize a liquid or gas cooling system or in which substantial quantities of fuel can flow from one part of the spacecraft to another.}.

Heat conduction is described by Fourier's law \cite{LL2002}:
\begin{equation}
\vec{q}=-k\nabla T,\label{eq:F1}
\end{equation}
where $\vec{q}$ is the heat flux (measured in units of power over area), $T$ is the temperature, and $k$ is the heat conduction coefficient. In the general case, $k$ is a tensorial quantity, but for homogeneous and isotropic materials, $k$ reduces to a scalar coefficient\footnote{A notable case of anisotropic conductivity where $k$ is tensorial is that of multilayer insulation.}. In general, $\vec{q}$, $k$, and $T$ are all functions of the coordinates $\vec{x}$ and time $t$.

Heat flux also obeys the energy conservation equation
\begin{equation}
\nabla\cdot\vec{q}=b-C_h\rho\frac{\partial T}{\partial t},\label{eq:F2}
\end{equation}
where $b$ is the volumetric heat release (measured in units of power density), $C_h$ is the material's specific heat, and $\rho$ is its density.

Equation (\ref{eq:F2}) has general applicability. In many cases of practical interest the internal heat sources are compact and pointlike, and their thermal power is known: for instance, instrumentation boxes within a spacecraft compartment. Denoting the thermal power of $n$ thermal sources as $B_i(t)$ ($i=1...n$), we can write
\begin{equation}
b(\vec{x},t)=\sum_{i=1}^nB_i(t)\delta^3(\vec{x}-\vec{x}_i),\label{eq:Bi}
\end{equation}
where $\vec{x}_i$ is the location of the $i$th heat source and $\delta$ is Dirac's delta function.

A system is in a {\em steady state} if its properties do not change with time. In particular, this means $\partial T/\partial t=0$, which leaves us with
\begin{equation}
\nabla\cdot\vec{q}=b.\label{eq:SS}
\end{equation}

The heat conducted to a surface element must equal the heat radiated by that surface element. Therefore, at the surface,
\begin{equation}
q=\vec{q}\cdot\vec{a},\label{eq:q}
\end{equation}
where $\vec{a}$ is the unit normal of the radiating surface element, and $q$ is the surface element's radiant intensity.

The radiant intensity, or energy flux, of a radiating surface is related to its temperature by the Stefan-Boltzmann law:
\begin{equation}
q(\vec{x},t)=\sigma\epsilon(\vec{x},t,T)T^4(\vec{x},t),
\label{eq:stefan-boltzmann}
\end{equation}
where $\sigma\simeq 5.67\times 10^{-8}$~Wm$^{-2}$K$^{-4}$ is the Stefan-Boltzmann constant, while the dimensionless coefficient $0\le\epsilon\le 1$ is a physical characteristic of the emitting surface. This coefficient can vary not only as a function of location and time, but also as a function of temperature.

We describe radiation by a four-dimensional stress-energy-momentum tensor that takes the form
\begin{equation}
T^{\mu\nu}=\begin{pmatrix}c^{-2}u&{\mathfrak p}\cr
{\mathfrak p}&\mathbb{P}\end{pmatrix},
\end{equation}
where $u$ is the energy density of the radiation field, ${\mathfrak p}$ is its momentum density, and $\mathbb{P}$ is the radiation pressure tensor. ($T^{\mu\nu}$ is, in fact, the stress-energy-momentum tensor of the electromagnetic field in a vacuum.)

The stress-energy-momentum tensor obeys the conservation equation
\begin{equation}
\nabla_\mu T^{\mu\nu}=0,
\end{equation}
where $\nabla_\mu$ denotes covariant differentiation with respect to the coordinate $x^\mu$. In three dimensions, this yields the following conservation equations:
\begin{eqnarray}
c^{-2}\dot{u}-\nabla\cdot{\mathfrak p}&=&0,\label{eq:dotu}\\
\dot{{\mathfrak p}}-\nabla\cdot\mathbb{P}&=&0,\label{eq:dotpi}
\end{eqnarray}
where the dot denotes differentiation with respect to time, i.e., $\dot{x}=\partial x/\partial t$.

The energy $E$ in a given volume $V$ is
\begin{equation}
E=\int udV
\end{equation}
by definition. The power, denoted by $Q$, is
\begin{equation}
Q=\frac{dE}{dt}=\int \dot{u}dV=c^2\int\nabla\cdot{\mathfrak p}dV,
\end{equation}
or, after applying Gauss's theorem,
\begin{equation}
Q=c^2\int{\mathfrak p}\cdot\vec{dA}=c^2\int{\mathfrak p}\cdot \vec{a}dA,\label{eq:dEdt1}
\end{equation}
where $A$ represents a surface enclosing the volume $V$ and $\vec{a}$ is the unit normal of surface element $\vec{dA}$ with area $dA$, such that $\vec{dA}=\vec{a} dA$.

Comparing with (\ref{eq:q}) and noting that $Q=\int qdA$, we obtain the relationship between radiant intensity and momentum density at the radiating surface:
\begin{equation}
q(\vec{x},t)=c^2{\mathfrak p}(\vec{x},t)\cdot\vec{a}.\label{eq:qpi}
\end{equation}

We describe the radiative flow of energy $E$ using the intensity\footnote{Some textbooks call the quantity $I$ the radiance, and its integral over a finite surface the (radiant) intensity.} $I$, which is the flow of energy across surface element $\vec{dA}$, in a time interval $dt$, in the solid angle $d\omega$ around direction $\vec{n}$:
\begin{equation}
E=\iiint I(\vec{x},t,\vec{n})\vec{n}\cdot\vec{dA}d\omega dt,
\end{equation}
or, after differentiating with respect to $t$,
\begin{equation}
Q=\frac{dE}{dt}=\iint I(\vec{x},t,\vec{n})\vec{n}\cdot\vec{dA}d\omega.\label{eq:dEdt2}
\end{equation}
Comparing with (\ref{eq:dEdt1}), we get
\begin{equation}
{\mathfrak p}=\frac{1}{c^2}\int I(\vec{x},t,\vec{n})\vec{n}d\omega.
\end{equation}
From (\ref{eq:qpi}), then, we obtain
\begin{equation}
q(\vec{x},t)=\int I(\vec{x},t,\vec{n})\vec{n}\cdot\vec{a}d\omega.
\end{equation}

A surface is {\it defined} as a diffuse or Lambertian emitter if the intensity $I$ does not depend on the direction of radiation emanating from a surface element. In this case, we can take $I$ outside the integral sign and write
\begin{equation}
q(\vec{x},t)=I(\vec{x},t)\int \vec{n}\cdot\vec{a}d\omega.
\label{eq:int1}
\end{equation}
The integral on the right-hand side should be evaluated over a hemispherical surface of unit radius centered around the surface element $\vec{dA}$. Parameterizing the integration surface using spherical coordinates $(r,\phi,\theta)$ (with $\phi=0$ at the north pole), we note that $d\omega=\sin\phi d\phi d\theta$ and $\vec{n}\cdot\vec{a}=\cos\phi$, and the integral reads
\begin{equation}
\int\vec{n}\cdot\vec{a}d\omega=\int_0^{2\pi}\int_0^{\pi/2}\cos\phi\sin\phi d\phi d\theta=\pi,
\end{equation}
thus
\begin{equation}
q(\vec{x},t)=\pi I(\vec{x},t).
\end{equation}

Radiation from a surface element $\vec{dA}=\vec{dA}_1$ that is intercepted by a second surface element $\vec{dA}_2$ at distance $r$ can be calculated by using, as the solid angle, $d\omega=r^{-2}\vec{n}\cdot\vec{dA}_2$ in (\ref{eq:dEdt2}):
\begin{eqnarray}
Q_{1\rightarrow 2}&=&\iint \frac{I(\vec{x},t,\vec{n})}{r^2}\vec{n}\cdot\vec{dA}_1\vec{n}\cdot\vec{dA}_2\nonumber\\
&=&\iint\frac{I(\vec{x},t,\vec{n})\cos\theta_1\cos\theta_2}{r^2}dA_1dA_2,
\end{eqnarray}
where $\theta_1$ and $\theta_2$, defined by $\cos\theta_1=\vec{n}\cdot\vec{dA}_1/dA_1$ and $\cos\theta_2=\vec{n}\cdot\vec{dA}_2/dA_2$, are the angles between the direction of heat radiation and the normals of the surface elements $\vec{dA}_1$ and $\vec{dA}_2$, respectively.

If both surfaces are Lambertian emitters, and we take into account heat flowing in both directions, the heat $Q_{12}$ exchanged between the two surfaces is
\begin{eqnarray}
Q_{12}&=&\iint\frac{(q_1-q_2)\cos\theta_1\cos\theta_2}{\pi r^2}dA_1dA_2\nonumber\\
&=&\iint\frac{(\epsilon_1T_1^4-\epsilon_2T_2^4)\cos\theta_1\cos\theta_2}{\pi r^2}dA_1dA_2.\label{eq:Q12}
\end{eqnarray}
These results describe the radiative exchange of energy. Next, we turn our attention to momentum exchange.

\section{Radiation pressure and the recoil force}
\label{sec:radpress}

The pressure tensor of radiation is written as \cite{MM1999}
\begin{equation}
\mathbb{P}(\vec{x},t)=\frac{1}{c}\int I(\vec{x},t,\vec{n})\vec{n}\vec{n}d\omega,
\end{equation}
where $\vec{u}\vec{v}$ is the dyadic product of two vectors $\vec{u}$ and $\vec{v}$. (This form of the pressure tensor coincides with the Maxwell stress tensor for plane or spherical electromagnetic waves.) For a Lambertian emitter, once again we can take $I$ outside the integral sign, yielding
\begin{equation}
\mathbb{P}(\vec{x},t)=\frac{1}{c}I(\vec{x},t)\int\vec{n}\vec{n}d\omega.
\end{equation}

The recoil force acting on the emitter will be of the same magnitude, but opposite in sign to the change in momentum in a given volume. It can be written as
\begin{equation}
\vec{F}(t)=-\int\dot{\mathfrak p}dV.
\end{equation}
After using (\ref{eq:dotpi}) and then applying Gauss's theorem, we get
\begin{equation}
\vec{F}(t)=-\int\nabla\cdot\mathbb{P}(\vec{x},t)dV=-\int\mathbb{P}(\vec{x},t)\cdot\vec{dA}.
\end{equation}
For a Lambertian emitter, this force then becomes \cite{ROS1990,ANT1992,VIG1994,DUH2006}
\begin{equation}
\vec{F}(t)=-\frac{1}{c}\int\left[I(\vec{x},t)\int\vec{n}\vec{n}\cdot\vec{a}d\omega\right]dA.
\end{equation}
The inner integral can be evaluated by making use of the identity $(\vec{a}\vec{b})\cdot\vec{c} = \vec{a}(\vec{b}\cdot\vec{c})$:
\begin{equation}
\int(\vec{n}\vec{n})\cdot\vec{a}d\omega=\int(\vec{a}\cdot\vec{n})\vec{n}d\omega.
\label{eq:int2}
\end{equation}
As with (\ref{eq:int1}), we integrate over a hemispherical surface of unit radius centered around the surface element $\vec{dA}$. We set up a spherical coordinate system $(r,\phi,\theta)$ such that $\phi=0$ corresponds to the north pole (that is, the direction of the surface normal $\vec{a}$), and we set up two additional basis vectors $\vec{b}(\phi=\pi/2,\theta=0)$ and $\vec{c}(\phi=\pi/2,\theta=\pi/2)$ such that $\vec{n}$ can be expressed as
\begin{eqnarray}
\vec{n}&=&(\vec{n}\cdot\vec{a})\vec{a}+(\vec{n}\cdot\vec{b})\vec{b}+(\vec{n}\cdot\vec{c})\vec{c}\nonumber\\
&=&\cos\phi\vec{a}+\sin\phi\cos\theta\vec{b}+\sin\phi\sin\theta\vec{c}.
\end{eqnarray}
To integrate (\ref{eq:int2}), we use $\vec{a}\cdot\vec{n}=\cos\phi$ to obtain
\begin{eqnarray}
&&\int\cos\phi[\cos\phi\vec{a}+\sin\phi\cos\theta\vec{b}+\sin\phi\sin\theta\vec{c}]d\omega\nonumber\\
&=&\int_0^{2\pi}\int_0^{\pi/2}\cos^2\phi\sin\phi\vec{a}d\phi d\theta=\frac{2\pi}{3}\vec{a}.
\end{eqnarray}
Therefore,
\begin{equation}
\vec{F}(t)=-\frac{2}{3}\frac{1}{c}\int q(\vec{x},t)\vec{dA},
\end{equation}
indicating that, in the Lambertian case, the radiation pressure is isotropic, and the radiation pressure tensor reduces to a scalar quantity:
\begin{equation}
\mathbb{P}(\vec{x},t)=\frac{2}{3}\frac{1}{c}q(\vec{x},t)\mathbb{I},
\label{eq:P}
\end{equation}
where $\mathbb{I}$ is the identity tensor.

As described previously, $q$ can be obtained by solving the heat conduction equations (\ref{eq:F1}) and (\ref{eq:F2}), along with the radiative heat transfer equation (\ref{eq:Q12}) and boundary conditions.

The surface density of the recoil force that corresponds to (\ref{eq:P}), acting on surface element $dA$ with unit normal $\vec{a}$, is
\begin{equation}
\vec{f}(\vec{x},t)=-\frac{2}{3}\frac{1}{c}q(\vec{x},t)\vec{a}.
\end{equation}
From this, the recoil force can be computed by integration\footnote{Knowing the recoil force surface density also allows us to compute the torque acting on the emitter. The torque surface density is $\vec{\tau}=(\vec{x}-\vec{x}_0)\times\vec{f}$, where $\vec{x}_0$ is the location of the the emitter's center-of-gravity. The total torque $\vec{T}$, then, can be written as
$$
\vec{T}(t)=\int(\vec{x}-\vec{x}_0)\times\vec{f}(\vec{x},t)dA=-\frac{2}{3}\frac{1}{c}\int q(\vec{x},t)(\vec{x}-\vec{x}_0)\times\vec{dA}.
$$}:
\begin{equation}
\vec{F}(t)=\int\vec{f}(\vec{x},t)dA=-\frac{2}{3}\frac{1}{c}\int q(\vec{x},t)\vec{dA}.
\label{eq:F}
\end{equation}
This result establishes the relationship we sought between radiated heat and the associated recoil force.

\section{The recoil force and its sources}
\label{sec:recoil}

Heat conduction $\vec{q}$ inside a heat emitting object and the radiant intensity $q$ at its exterior surfaces can be obtained by solving, in the steady state, Eqs. (\ref{eq:SS}) and (\ref{eq:q}) for the unknown functions $\vec{q}$ and $q$ using the known functions $\vec{a}$ (representing the emitter's geometry) and $b$ (the volumetric heat release inside the emitter), along with appropriate boundary conditions (e.g., sky temperature).
Given constant boundary conditions and an unchanging geometry, the solution for $q(\vec{x},t)$ can be obtained in terms of $b(\vec{x},t)$. The recoil force $\vec{F}(t)$, which is a functional of $q(\vec{x},t)$ as per (\ref{eq:F}), can therefore be expressed as a functional of $b(\vec{x},t)$:
\begin{equation}
\vec{F}(t)=\vec{F}[b(\vec{x},t)].
\end{equation}
In many cases, $b$ can be represented by discrete, compact heat sources in accordance with (\ref{eq:Bi}). As we noted, this is the case, in particular for spacecraft containing instrumentation boxes within its compartment, with telemetered power readings available for each. In this case, we can write $\vec{F}$ as a functional of the $n$ functions $B_i(t)$:
\begin{equation}
\vec{F}(t)=\vec{F}[B_1(t),B_2(t),...,B_n(t)].
\label{eq:FB}
\end{equation}

The magnitude and direction of the recoil force are both functions of time. However, if the emitting object is rotating, and its rate of rotation is sufficiently high compared to the rate of change of the recoil force, the time average vector components of the recoil force that lie in the plane of rotation will be negligible. The residual recoil force will always be perpendicular to the plane of rotation, i.e., parallel to the rotating object's spin axis.

To see this, we write the recoil force in the form
\begin{equation}
\vec{F}(t)=\vec{F}_\parallel(t)+\mathbb{R}(\omega t)\cdot\vec{F}_\perp(t),
\end{equation}
where
\begin{equation}
\vec{F}_\parallel(t)=\big(\vec{F}(t)\cdot\vec{s}\big)\vec{s}
\end{equation}
is the component of $\vec{F}(t)$ parallel with the spin axis represented by the unit vector $\vec{s}$ (which, we assume, remains constant in time), and
\begin{equation}
\vec{F}_\perp(t)=\mathbb{R}^{-1}(\omega t_0)\big(\vec{F}(t)-\vec{F}_\parallel(t)\big)
\end{equation}
is a perpendicular component of $\vec{F}(t)$ in a corotating reference frame, while $\mathbb{R}(\phi)$ is a tensor representing a rotation by the angle $\phi$.

Ignoring forces other than the recoil force, the position of the rotating object as a function of time can be calculated as
\begin{equation}
\vec{x}(t)=\vec{x}(t_0)+\dot{\vec{x}}(t_0)(t-t_0)+\iint\frac{1}{m}\vec{F}(t)d^2t,
\end{equation}
where $\vec{x}(t_0)$ is the position, $\dot{\vec{x}}(t_0)$ the velocity of the object at some time $t=t_0$. We assume that the object's mass, $m$, remains constant in time. We denote the displacement of the object as a result of the recoil force as $\Delta\vec{x}=\vec{x}(t)-\vec{x}_0-\vec{v}_0t$. Therefore,
\begin{equation}
\Delta\vec{x}=\frac{1}{m}\iint\big(\vec{F}_\parallel(t)+\mathbb{R}(\omega t)\cdot\vec{F}_\perp(t)\big)d^2t.
\end{equation}
The displacement due to $\vec{F}_\parallel$ is entirely along the spin axis. To calculate the displacement due to the perpendicular component, we assume that it is well approximated by a linear function of time:
\begin{equation}
\vec{F}_\perp(t)=\vec{F}_0+\dot{\vec{F}}_\perp t,
\end{equation}
where $\dot{\vec{F}}_\perp$ is constant. Thereafter, noting that $\int\mathbb{R}(\omega t)dt=\omega^{-1}\mathbb{R}(\omega t-\pi/2)$, we calculate the double integral for $n$ full revolutions over a time period $\Delta t=2\pi n/\omega$ and obtain

\begin{equation}
\Delta\vec{x}_\perp=\frac{1}{\omega^2m}\mathbb{R}(\omega t_0-\pi)\cdot\dot{\vec{F}}_\perp\Delta t,
\end{equation}
describing an arithmetic spiral. In three dimensions, then, the motion of the object is described by a helix of widening radius around the object's original axis of rotation. The growth of the width of the helix is governed by $\dot{\vec{F}}_\perp$. If $\vec{F}_\perp$ remains nearly constant in time, $\dot{\vec{F}}_\perp\simeq 0$ and the object's spin axis returns to its original position after every full revolution with no cumulative displacement in the perpendicular direction. Therefore, ignoring a constant $\vec{F}_\perp$ introduces only a small periodic error and no cumulative error in the calculations, and ignoring $\vec{F}_\perp$ in the case when $\dot{\vec{F}}_\perp$ is nonvanishing but small and approximately constant only introduces a small cumulative error. This means that in the case of a spinning object, the recoil force (\ref{eq:FB}) can be expressed as
\begin{equation}
\vec{F}(t)=F_\parallel[B_1(t),B_2(t),...,B_n(t)]\vec{s}.
\label{eq:FBi}
\end{equation}

Without loss of generality, $F_\parallel$ can be expanded in the form of a Taylor series in the $B_i$. For the purposes of the present paper, it is sufficient to keep only linear terms. With this in mind, and noting that $F_\parallel(0,...,0)=0$, we can write
\begin{equation}
F_\parallel(t)=\frac{1}{c}\sum_{i=1}^n\xi_iB_i(t),
\label{eq:Fxi}
\end{equation}
where the dimensionless coefficients $\xi_i$ are given by
\begin{equation}
\xi_i=c\frac{\partial F_\parallel[B_i(t)]}{\partial B_i(t)}.
\label{eq:xii}
\end{equation}
The factors $\xi_i$ are determined by the geometry and optical properties of the emitter, and are expected to remain constant so long as the emitter's geometry and optical properties do not change.

For any given heat source, the principle of conservation of energy dictates that $-1\le\xi_i\le 1$. The coefficient is zero if heat from that particular source is emitted isotropically, resulting in no net recoil force. Therefore, these factors determine the efficiency with which each heat source contributes to the object's acceleration.

\section{Application and error analysis}
\label{sec:errors}

The formalism that we obtained can be employed in a direct calculation of the recoil force using conventional numerical methods of heat transfer. Together, Eqs.~(\ref{eq:F1}) and (\ref{eq:F2}) are two first-order differential equations in two unknown functions $\vec{q}$ and $T$ that describe heat conduction inside materials, while (\ref{eq:Q12}) represents additional constraint equations describing radiatively coupled surface elements.

A specific solution can be obtained if the material properties (represented by $k$ and $\epsilon$, as well as $C_h$ and $\rho$) and heat sources (represented by $b$) are known, and appropriate boundary conditions are given.

One such boundary condition is the steady-state condition $\partial T/\partial t=0$, in which case we replace (\ref{eq:F2}) with (\ref{eq:SS}). For a heat emitting object situated in empty space, another boundary condition can be specified in the form of the sky temperature (i.e., the microwave background radiation temperature) to which the object's exterior surfaces are radiatively coupled. A practical difficulty arises if the object has facing surfaces (i.e., surfaces that are radiatively coupled to one another) but these situations are dealt with easily using standard finite element or ray-tracing numerical codes.

It should be noted that in this case, a solution is fully specified when the volumetric heat release $b(\vec{x},t)$ is known, even as no temperature values inside the emitting object are given. When spacecraft telemetry provides both power and temperature measurements, these data together represent a redundant data set that can be used to verify and validate thermal models. Laborious but, in principle, straightforward application of these equations can lead to a temperature map of the exterior surfaces of an emitter. When this temperature map is known, the recoil force can be computed directly using Eq.~(\ref{eq:F}).

An alternative to evaluating the vector-valued integral (\ref{eq:F}) along the nontrivial exterior geometry of the emitter is the use of a control volume technique \cite{MEXPROC2007}. The anisotropy of thermal emissions can be determined by surrounding the object with an infinite control volume, which is approximated by a sufficiently large fictitious spherical surface centered around the emitter that is used to intercept all radiated heat coming from the emitter (see discussion of a similar approach involving pixel arrays in \cite{ZIE2005}). The recoil force is computed by evaluating the integral (\ref{eq:F}) along the surface of this sphere. These calculations can be carried out using standard thermal modeling tools\footnote{For instance, Thermal Desktop, the Thermal Radiation Analysis System (TRASYS) and Systems Improved Numerical Differencing Analyzer (SINDA) \cite{ANT1992,MEXPROC2007}.}, that have been used successfully for the design and operations of many missions at the Jet Propulsion Laboratory (JPL).

How accurately can the thermal recoil force be determined? It always has been recognized that accurate computation of this force is a difficult task. Computing the total amount of heat emitted by a spacecraft is straightforward: if the thermal power of internal heat sources is known and the spacecraft is in a steady state, the amount of heat it radiates must equal the generated heat.

The recoil force, however, depends not only on the total amount of heat radiated by the spacecraft, but on the differences in heat radiated in different directions. These, in turn, are calculated using detailed knowledge of the spacecraft's geometry and material properties, which may be poorly known.

Considering the force model (\ref{eq:Fxi}), we note that once a comprehensive numerical model is available, it needs to be evaluated only a modest number of times in order to obtain the efficiency factors $\xi_i$ through Eq.~(\ref{eq:xii}). Additional evaluations can be used to determine the error of this homogeneous and linear approximation, which we denote with $\sigma_\mathrm{model}$. If we assume that sources of error are independent and not correlated, the error of the recoil force estimate (\ref{eq:Fxi}) can then be written as
\begin{equation}
\sigma_F^2(t)=\frac{1}{c^2}\sum_{i=1}^n\big(\sigma_{\xi_i}^2B^2_i(t)+\xi^2_i\sigma_{B_i}^2(t)\big)+\sigma_\mathrm{model}^2,
\end{equation}
where uncertainties in the knowledge of $B_i(t)$ are represented by the standard deviations $\sigma_{B_i}(t)$, and uncertainties in the calculation of $\xi_i$ are represented by the standard deviations $\sigma_{\xi_i}$.

For spacecraft, the values of $B_i(t)$ are either measured values from telemetry, or nominal values from design documentation. In the former case, $\sigma_{B_i}(t)$ can be obtained by considering sensor sensitivity and telemetry resolution. In the latter case, uncertainties may be available from design documentation, or may be inferred, e.g., by comparing nominal power consumption with measured values of available electrical power.

The values of the $\sigma_{\xi_i}$ are the most difficult to estimate. These values must be developed on a case-by-case basis, accounting for uncertainties in the knowledge of the spacecraft's geometry, material properties, and physical configuration, as well as the effects of aging on these.

\section{Orbit determination}
\label{sec:orbit}

The position of a distant spacecraft is rarely observed directly. Instead, the spacecraft's position is inferred from radio-metric observables, notably radio-metric Doppler and range measurements. The expected values of these measurements can be computed if the spacecraft's position and velocity are known. Not including general relativistic corrections \cite{MOYER2000,MG2005}, the spacecraft's equation of motion can be written in the general form
\begin{equation}
\ddot{\vec{r}}=\sum_{j=1}^lGM_j\frac{\vec{r}_j-\vec{r}}{|\vec{r}_j-\vec{r}|^3}+\frac{1}{m}\sum_k\vec{F}_k,
\label{eq:r}
\end{equation}
where $\vec{r}$ is the position of the spacecraft, $G$ is Newton's gravitational constant, $M_j$ are the masses and $\vec{r}_j$ are the positions of bodies that influence the spacecraft's position gravitationally (these would typically include all major solar system bodies, and any smaller bodies that are sufficiently near the spacecraft to affect its orbit), $m$ is the spacecraft's mass, and $\vec{F}_k$ are any nongravitational forces acting on the spacecraft.

Once the orbit of the spacecraft is known, the expected values of radio-metric observables can be calculated by taking into account signal propagation in the solar system environment, atmospheric effects, antenna locations, the relative motion of antennae and spacecraft, and other effects such as the spacecraft's spin that can influence the signal.

Equation~(\ref{eq:r}) is a second-order differential equation in $\vec{r}$. To obtain a specific solution of such an equation, one needs suitable initial conditions, which can be specified in the form of the initial state vector that consists of $\vec{r}(t_0)$ and $\dot{\vec{r}}(t_0)$ at some time $t=t_0$. The purpose of the orbit determination effort is to find initial conditions for which the difference between computed and observed radio-metric values, i.e., the residual, is minimized.

The small forces model in (\ref{eq:r}) may be parameterized. If the parameter values are unknown, the orbit determination exercise can be used to find these along with the initial state vector.

As an example, solar pressure may be represented as a force acting on the spacecraft, modeling the spacecraft's geometry, with given solar absorptance, specular, and diffuse reflectivity coefficients of its surfaces. These coefficients may not be known in advance accurately, but they can be determined along with the orbital initial conditions through minimizing the residuals of radio-metric observables by simultaneous adjustment of the initial state vector and the solar pressure coefficients.

Precision orbit determination requires the repeated evaluation of (\ref{eq:r}) a large number of times. Therefore, it is essential that the forces $\vec{F}_k$ can be computed with as little computational overhead as possible. In the case of the thermal recoil force, this precludes the possibility of evaluating a comprehensive thermal model multiple times at every point along the spacecraft's orbit. However, after the coefficients $\xi_i$ have been determined by evaluating the comprehensive thermal model a limited number of times, the thermal recoil force (\ref{eq:FBi}) can be incorporated easily into the orbit determination process.

In a more innovative approach, we can treat the coefficients $\xi_i$ as unknown parameters, and use the orbit determination program to determine their values, along with the initial state vector and other parameters. This approach is especially notable because it is applicable even when no detailed thermal model for a spacecraft is available. When a detailed thermal model is present, agreement between the two methods validates that model and the hypothesis that no additional forces of unknown origin act on the spacecraft. Conversely, if the two calculations are in significant disagreement, that can be a strong indication that an anomalous force of unknown origin is present.

So long as the amount of heat generated by on-board components is known and the spacecraft's geometric and optical properties are not changing with time, one can make the assumption that the on-board generated recoil force is well modeled by (\ref{eq:Fxi}) and that no other unknown forces affect the spacecraft, and then proceed to verify this model and at the same time determine the values of $\xi_i$ by fitting to the orbital data. Conversely, even when a detailed thermal model is available, independent determination of the $\xi_i$ from orbital data alone offers a robust way to verify the results of the thermal model and help confirm or reject any hypotheses concerning the presence of additional forces of unknown origin.

\section{The case of the Pioneer anomaly}
\label{sec:application}

We have applied the methods that we developed in the previous sections to the study of the Pioneer anomaly \cite{JPL2002}.

Pioneer 10 and 11, launched in March 1972 and April 1973, respectively, were the first man-made objects to travel to the outer regions of the solar system and beyond. After flying by Jupiter (and, in the case of Pioneer~11, Saturn), the spacecraft continued on hyperbolic escape trajectories, while they were being tracked by NASA's Deep Space Network system of radio tracking stations. Pioneer~11 remained operational until 1995, although precision navigation of this spacecraft ended in 1990 due to on-board failures. Pioneer~10 was operating as late as 2003, and precision Doppler measurements were received from this spacecraft until the end of its mission\footnote{Only Doppler; Pioneer~10 and 11 had no range observable.}.

The Pioneer spacecraft \cite{PC202} were spin stabilized. Their spin axis, coinciding with the antenna axis, was pointed towards the Earth to ensure continuous communication. Spin stabilization meant that trajectory correction maneuvers were infrequent; most of the time late in their missions, Pioneer 10 and 11 were flying undisturbed. Because of this, the twin Pioneers were considered a reliable platform for precision gravitational measurements, searching for a planet beyond Pluto, and for gravitational waves.

While neither ``planet X'' nor gravitational waves were detected, there remained an unexplained residual between calculated and observed Doppler data \cite{JPL1998,JPL2002}. This residual can be eliminated by assuming that a constant acceleration of unknown origin pushes the spacecraft towards the Sun. The magnitude of this acceleration is
\begin{equation}
a_P=(8.74\pm 1.33)\times 10^{-10}~\mathrm{m}/\mathrm{s}^2.\label{eq:aP}
\end{equation}

\begin{figure}
\includegraphics[width=\linewidth]{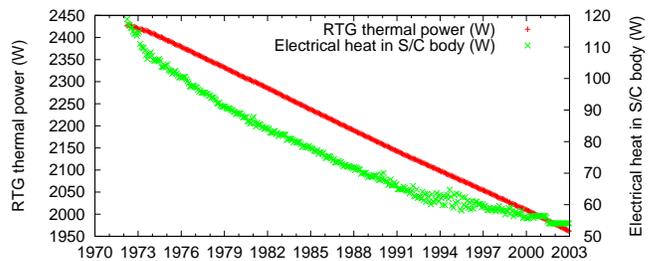}\\
\caption{RTG heat (red +) and electrical heat (green $\times$), measured in W, on board Pioneer 10, from telemetry \cite{MDR2005,MEXPROC2007}.}
\label{fig:power}
\end{figure}

A possible origin of this approximately constant acceleration could be a thermal recoil force. Electrical power on board the Pioneer spacecraft was generated by low-efficiency RTGs that produced waste heat in excess of 2~kW \cite{MDR2005} throughout the mission. Electrical equipment on board the spacecraft produced an additional $\sim 100$~W of heat (see Fig.~\ref{fig:power}). This heat was radiated away by the spacecraft in a complex pattern, as determined by the spacecraft's geometry and material properties. A small anisotropy, less than 2\% in magnitude, would be sufficient to provide the necessary force to yield the anomalous acceleration (\ref{eq:aP}).

The RTGs are mounted on the Pioneer spacecraft at the end of booms that are approximately 3~m in length. The RTGs are compact objects. This suggests that the recoil force due to RTG heat may be modeled accurately as a homogeneous linear function of the RTG heat, in accordance with the discussion in Sec.~\ref{sec:recoil}. Further, although each spacecraft has four RTGs, they are mounted symmetrically and their temporal behavior is nearly identical: therefore, they can be treated as a single heat source.

Electrical heat is produced inside the spacecraft body. Most of this body is covered by multilayer thermal insulation, resulting in small exterior temperature differences. This implies that the recoil force due to electrically generated heat can also be a homogeneous linear function of the electrical heat, and further, the possibility that any particular distribution of heat sources inside the spacecraft body can be neglected, only their total thermal output must be considered.

Additional heat sources on board the spacecraft include 12 small (1~W) radioisotope heater units (RHUs) and the propulsion system. The former can be ignored due to their geometry; most of the heat produced by the RHUs is radiated in a direction perpendicular to the spin axis\footnote{Private communication from Jim Moses, TRW retiree.}. Heat generated by the propulsion system, in turn, can be ignored as these events are transient, and any thermal recoil force due to propulsion system heating is masked by uncertainties in the modeling of the maneuvers.

Further, as the Pioneer spacecraft are spinning, we only need to consider the recoil force in the spin axis direction, according to Sec.~\ref{sec:recoil}.

Temperatures inside the spacecraft are nearly constant, changing only on the times cale of years. Therefore, the spacecraft is accurately described by a steady-state model.

Using recently recovered documentation \cite{MDR2005}, a highly detailed finite element model incorporating $\sim$3000 nodes and 2600 plate elements, using $3.4\times 10^6$ radiation conductors and $\sim$7000 linear conductors has been constructed \cite{MEXPROC2007}. Temperatures and power readings from telemetry were used as a redundant data set to establish boundary conditions. Analysis of this model is presently under way and results will be reported when they become available.

This model will also be used to validate the assumptions leading to (\ref{eq:Fxi}), notably by verifying that the resulting recoil force is indeed a homogeneous linear functional of the RTG and electrical heat:
\begin{equation}
\vec{F}(t)=\frac{1}{c}(\xi_rB_r(t)+\xi_eB_e(t))\vec{s},\label{eq:Fer}
\end{equation}
where $\xi_r$ and $\xi_e$ are the efficiency factors associated with RTG thermal power $B_r(t)$ and electrical power $B_e(t)$. As before, $\vec{s}$ is a unit vector pointing in the spin axis direction.

The discrepancy between the linear model (\ref{eq:Fer}) and a comprehensive thermal model can be evaluated to yield an estimate of $\sigma_\mathrm{model}$. We have not yet computed the $\sigma_\mathrm{model}$ numerically.

The value of $B_e(t)$ is available from telemetry (Fig.~\ref{fig:power}). Uncertainties in this value are due primarily to two factors. First, telemetry has limited resolution (analog sensor readings are telemetered after conversion to 6-bit binary values). Second, for some instruments on board, only their nominal power consumption is known, telemetry provides only their on/off state, not their actual power level. Taking all these uncertainties into account, we calculate
\begin{equation}
\sigma_{B_e}=1.8~\mathrm{W}.
\end{equation}

The total power of the RTGs is known precisely from prelaunch documentation. The physics of the radioactive decay of the $^{238}$Pu fuel is well understood. The amount of power removed from the RTGs in the form of electrical energy is telemetered to the ground, and thus the time dependence of the RTG thermal power $B_r(t)$ is known. The primary source of uncertainty in the calculation of $B_r(t)$ is the limited resolution of this telemetry. This uncertainty is calculated as
\begin{equation}
\sigma_{B_r}=1.1~\mathrm{W}.
\end{equation}

We note that $\sigma_{B_e}$ and $\sigma_{B_r}$ are anticorrelated; if the electrical heat is overestimated, the RTG heat is underestimated using the same telemetry, and vice versa. Treating the two sources of error as uncorrelated, therefore, yields a conservative error estimate.

Additional temperature information is available in the telemetry stream, measured by sensors located at various points around the spacecraft. These temperature readings offer redundant information about the thermal state of the spacecraft that can be used to verify and validate thermal models.

The effort to develop a comprehensive thermal model of the Pioneer 10 and 11 spacecraft is on-going. This effort is expected to determine an estimate for $\xi_e$ and $\xi_r$. A naive analytical model of the spacecraft, verified by a simple ray-tracing computational model, suggests that the values must be approximately
\begin{eqnarray}
\xi_e\simeq 0.36,&&\xi_r\simeq 0.010,\label{eq:xi}
\end{eqnarray}
albeit the relative error on these figures may be a high as 100\% or more, due to the simplicity of the models that were used to obtain them. Nevertheless, we use these figures as typical figures in the present analysis.

We have recently developed a precision orbit determination program \cite{TOTH2008} that can process Pioneer 10 and 11 Doppler data. It uses the latest JPL ephemerides (DE-414) to determine the position of solar system bodies. The program models spacecraft orbits using relativistic equations of motion. It accurately models signal propagation by taking into account, for instance, the Shapiro time delay and effects of the troposphere and solar plasma on the radio signal. This program has been used successfully to confirm the existence of the Pioneer anomaly \cite{TOTH2008}. The program also has the capability to utilize spacecraft telemetry and model on-board generated thermal recoil forces (\ref{eq:Fxi}).

An effort to recover all available Pioneer 10 and 11 data is presently on-going \cite{MDR2005,MEXPROC2007}. Before we apply our method to this soon complete data set, it was essential to demonstrate the viability of our method. Notably, we would like to know if it is possible, in principle, to distinguish between a constant sunward acceleration and a thermal recoil force. For this purpose, we built simulated sets of Pioneer 10 Doppler data. In one particular simulation, we used the following fictitious values of a constant acceleration term $a_0$ and thermal coefficients $\xi_e$ and $\xi_r$:

\begin{eqnarray}
a_0&=&2\times 10^{-10}~\mathrm{m}/\mathrm{s}^2,\\
\xi_e&=&0.3,\\
\xi_r&=&0.015,
\end{eqnarray}
consistent with (\ref{eq:xi}).

Furthermore, the simulation utilized actual Pioneer 10 telemetry to model the internal heat of the spacecraft. The simulated data set ran from 1987 to 1998 and contained 13,534 Doppler data points. To make the simulation realistic, Gaussian random noise with $\sigma=5$~mHz was added to the Doppler data. Additionally, a sinusoidal diurnal term and a sinusoidal annual term, both with a peak-to-peak amplitude of 10~mHz, were added to the signal, to simulate possible mismodeling, by effects such as those of the atmosphere on the signal and of the dynamics of the solar system.

\begin{figure}
\includegraphics[width=\linewidth]{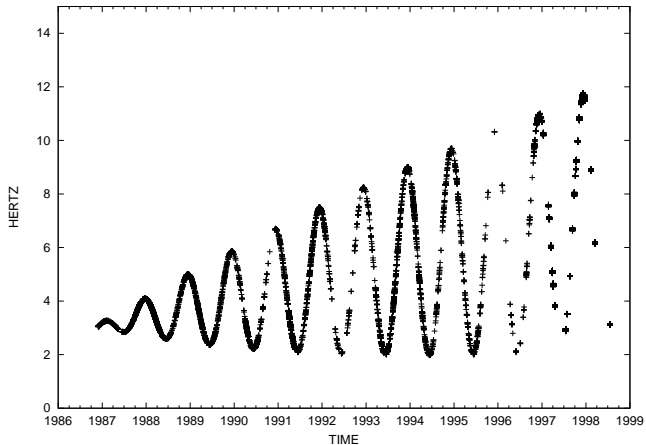}\\
\caption{Simulated Pioneer 10 Doppler prefit residuals.}
\label{fig:qinitial}
\end{figure}

Position values rounded to the nearest 1000~km and velocities rounded to the nearest m/s were used as the initial state vector (this is the typical magnitude of error we observe when we use Pioneer ephemeris data from JPL Horizons On-Line Ephemeris System\footnote{\url{http://ssd.jpl.nasa.gov/?horizons}.} for initial conditions.) We used initial values of $a_0=1\times 10^{-10}$~m/s$^2$, $\xi_e=0$, and $\xi_r=0$. Though we have the capability to deal with maneuvers in the actual data, for the purposes of this exercise we did not simulate maneuvers. The resulting prefit residuals are shown in Fig.~\ref{fig:qinitial}.

\begin{figure}
\includegraphics[width=\linewidth]{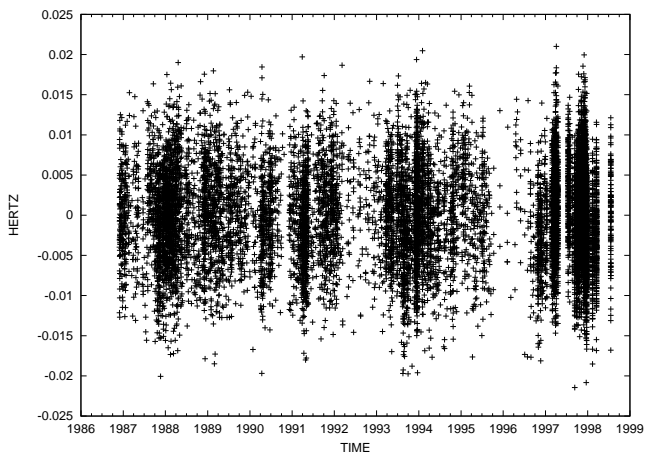}\\
\caption{Post-fit residuals of the simulated Pioneer 10 data set after the values of $a_0$, $\xi_e$, and $\xi_r$ along with the initial state vector were fitted successfully.}
\label{fig:qresult}
\end{figure}

The goal of this simulation was to demonstrate that even in the presence of noisy data, a constant acceleration term and acceleration due to thermal radiation can be clearly distinguished. Despite the presence of noise, our orbit determination algorithm successfully recovered the values of
\begin{eqnarray}
a_0&=&(2.1107\pm 0.0170)\times 10^{-10}~\mathrm{m}/\mathrm{s}^2,\\
\xi_e&=&0.292~50\pm 0.002~54,\\
\xi_r&=&0.014~856\pm 0.000~081.
\end{eqnarray}
The post-fit residuals are shown in Fig.~\ref{fig:qresult}. The root mean square residual of this solution is 5.84~mHz, which corresponds to the noise that was added to this simulated data set.

\begin{table}
\caption{Covariance matrix elements for $a_0$, $\xi_r$ and $\xi_e$.}
\label{tb:subcov}
\begin{tabular}{cccc}\hline\hline
&$a_0$&$\xi_r$&$\xi_e$\\
\hline
$a_0$&$~~2.91\times 10^{-24}$&$-2.15\times 10^{-19}$&$-2.95\times 10^{-15}$\\
$\xi_r$&$-2.15\times 10^{-19}$&$~~6.51\times 10^{-9~}$&$-1.20\times 10^{-7~}$\\
$\xi_e$&$-2.95\times 10^{-15}$&$-1.20\times 10^{-7~}$&$~~6.45\times 10^{-6~}$\\
\hline\hline
\end{tabular}
\end{table}

Furthermore, the cross-correlation between $a_0$, $\xi_e$, and $\xi_r$ remains small. This can be seen by visual inspection of the relevant elements of the covariance matrix, shown in Table~\ref{tb:subcov}. We note that, after normalizing using the values of $a_0$, $\xi_r$, and $\xi_e$, the diagonal elements of the covariance matrix dominate.

These results indicate that the approach we presented is feasible. It is possible, in principle, to distinguish a constant acceleration from time-varying acceleration due to thermal radiation using Doppler data alone, even when the data has a moderate amount of noise. Nonetheless, it is imperative to reduce the noise in the data as much as possible, for example by carefully modeling small effects such as those of the atmosphere and solar plasma on the spacecraft's radio signal, or small accelerations due to fuel leaks and maneuver uncertainties.

Once an improved thermal model becomes available, it can be used to verify the linear hypothesis expressed in (\ref{eq:Fxi}), which forms the basis of the approach we present here. The thermal model may also be used to quantify the error margins on $\xi_e$ and $\xi_r$. On the other hand, analysis of recently recovered Doppler data can confirm if the orbital behavior of the Pioneer spacecraft remained consistent throughout their missions, and may also help reduce the error margins on any residual acceleration that remains after accounting for the thermal recoil force.

\section{Conclusions}
\label{sec:end}

An object that emits heat experiences a recoil force due to radiation pressure. In this paper, we developed the basic equations that can be used to estimate the magnitude of this recoil force, and relate the recoil force to the amount of heat produced internally. We have been able to show how, under specific circumstances, the recoil force can be modeled as an homogeneous linear function of the power of discrete internal power sources. When this approach is applicable, the linear relationship can be readily incorporated into orbit determination efforts.

To analyze the trajectory of Pioneer 10 and 11, we developed orbit determination software that estimates the thermal recoil force acting on the spacecraft. Our software uses telemetry information as it calculates the thermal power of on-board heat sources as functions of time.

A comprehensive thermal model, presently under development, will allow us to verify the key assumptions behind our modeling, most notably the assertion that the thermal recoil force is accurately modeled as a linear, homogeneous function of electrical heat and heat from the radioisotope thermoelectric generators.

Using a simulated Doppler data set and actual Pioneer 10 telemetry, we demonstrated that it is possible in principle to distinguish acceleration due thermal radiation from a constant sunward acceleration term.

Newly recovered Doppler data are now available as a result of an extensive data recovery effort \cite{MDR2005,MEXPROC2007}. This will allow us to extend our analysis, and verify whether or not the thermal recoil force can account for the anomalous acceleration of Pioneer~10 and 11. These results will be published elsewhere when they become available.

We emphasize that the approach presented here, notably the direct utilization of flight telemetry in precision spacecraft navigation codes, has never been attempted before. The approach we describe is applicable not only to the case of Pioneer~10 and 11, but also to the case of present and future spacecraft. One mission in particular that may benefit from this approach is New Horizons, on its way towards an encounter with Pluto in 2015. While presently not used for gravitational research, such investigations could be conducted during its multiyear cruise. If such an investigation is undertaken, it will require accurate estimates of the thermal recoil force due to the waste heat produced by New Horizons' RTG and electrical equipment.

~\par

\section*{ACKNOWLEDGMENTS}

This work was initiated during our visit to the Perimeter Institute for Theoretical Physics, Waterloo, Canada. We thank John Moffat for his hospitality and support.
Part of this work was carried out at the International Space Science Institute (ISSI), Bern, Switzerland. We thank Roger M. Bonnet, Vittorio Manno, Brigitte Fasler, Saliba F. Saliba and members of the ISSI staff.
We would like to express our gratitude to Gary Kinsella of JPL who benefited us with his insightful comments and suggestions regarding thermal modeling of the Pioneers. We thank Craig B. Markwardt and Louis K. Scheffer for helpful discussions and Larry Kellogg and Dave Lozier for valuable advice.
The work of S.G.T. was carried out at the Jet Propulsion Laboratory, California Institute of Technology, under a contract with the National Aeronautics and Space Administration.

\bibliography{refs}

\bibliographystyle{apsrev}

\end{document}